\documentclass[aps,pra,preprint,superscriptaddress]{revtex4}

\usepackage{graphicx}
\usepackage{amsmath}
\usepackage{amssymb}
\usepackage{tabularx}

\begin{document}


\title{First storage of ion beams in the Double Electrostatic Ion-Ring Experiment - DESIREE}

\author{}
\author{H.~T.~Schmidt}
\affiliation{Department of Physics, Stockholm University, SE-10691 Stockholm, Sweden}
\author{R.~D.~Thomas}
\affiliation{Department of Physics, Stockholm University, SE-10691 Stockholm, Sweden}
\author{M.~Gatchell}
\affiliation{Department of Physics, Stockholm University, SE-10691 Stockholm, Sweden}
\author{S.~Ros\'en}
\affiliation{Department of Physics, Stockholm University, SE-10691 Stockholm, Sweden}
\author{P.~Reinhed}
\affiliation{Department of Physics, Stockholm University, SE-10691 Stockholm, Sweden}
\author{P.~L\"ofgren}
\affiliation{Department of Physics, Stockholm University, SE-10691 Stockholm, Sweden}
\author{L.~Br\"annholm}
\affiliation{Department of Physics, Stockholm University, SE-10691 Stockholm, Sweden}
\author{M.~Blom}
\affiliation{Department of Physics, Stockholm University, SE-10691 Stockholm, Sweden}
\author{M.~Bj\"orkhage}
\affiliation{Department of Physics, Stockholm University, SE-10691 Stockholm, Sweden}
\author{E.~B\"ackstr\"om}
\affiliation{Department of Physics, Stockholm University, SE-10691 Stockholm, Sweden}
\author{J.~D.~Alexander}
\affiliation{Department of Physics, Stockholm University, SE-10691 Stockholm, Sweden}
\author{S.~Leontein}
\affiliation{Department of Physics, Stockholm University, SE-10691 Stockholm, Sweden}
\author{D.~Hanstorp}
\affiliation{Department of Physics, University of Gothenburg, SE-41296 Gothenburg, Sweden}
\author{H.~Zettergren}
\affiliation{Department of Physics, Stockholm University, SE-10691 Stockholm, Sweden}
\author{L.~Liljeby}
\affiliation{Department of Physics, Stockholm University, SE-10691 Stockholm, Sweden}
\author{A.~K\"allberg}
\affiliation{Department of Physics, Stockholm University, SE-10691 Stockholm, Sweden}
\author{A.~Simonsson}
\affiliation{Department of Physics, Stockholm University, SE-10691 Stockholm, Sweden}
\author{F.~Hellberg}
\affiliation{Department of Physics, Stockholm University, SE-10691 Stockholm, Sweden}
\author{S.~Mannervik}
\affiliation{Department of Physics, Stockholm University, SE-10691 Stockholm, Sweden}
\author{M.~Larsson}
\affiliation{Department of Physics, Stockholm University, SE-10691 Stockholm, Sweden}
\author{W.~D.~Geppert}
\affiliation{Department of Physics, Stockholm University, SE-10691 Stockholm, Sweden}
\author{K.~G.~Rensfelt}
\affiliation{Department of Physics, Stockholm University, SE-10691 Stockholm, Sweden}
\author{H.~Danared}
\affiliation{European Spallation Source ESS AB, Box 176, SE-221 00 Lund, Sweden}
\affiliation{Department of Physics, Stockholm University, SE-10691 Stockholm, Sweden}
\author{A.~Pa\'al}
\affiliation{Department of Physics, Stockholm University, SE-10691 Stockholm, Sweden}
\author{M.~Masuda}
\affiliation{Department of Physics, Stockholm University, SE-10691 Stockholm, Sweden}
\author{P.~Halld\'en}
\affiliation{Department of Physics, Stockholm University, SE-10691 Stockholm, Sweden}
\author{G.~Andler}
\affiliation{Department of Physics, Stockholm University, SE-10691 Stockholm, Sweden}
\author{M.~H.~Stockett}
\affiliation{Department of Physics, Stockholm University, SE-10691 Stockholm, Sweden}
\author{T.~Chen}
\affiliation{Department of Physics, Stockholm University, SE-10691 Stockholm, Sweden}
\author{G.~K\"allersj\"o}
\affiliation{Department of Physics, Stockholm University, SE-10691 Stockholm, Sweden}
\author{J.~Weimer}
\affiliation{Department of Physics, Stockholm University, SE-10691 Stockholm, Sweden}
\author{K.~Hansen}
\affiliation{Department of Physics, University of Gothenburg, SE-41296 Gothenburg, Sweden}
\author{H.~Hartman}
\affiliation{Applied Mathematics and Material Science, Malm\"o University, SE-20506 Malm\"o, Sweden}
\affiliation{Lund Observatory, Lund University, SE-22100 Lund, Sweden}
\author{H.~Cederquist}
\affiliation{Department of Physics, Stockholm University, SE-10691 Stockholm, Sweden}
\email{schmidt@fysik.su.se}


\date{\today}

\begin{abstract}
We report on the first storage of ion beams in the Double ElectroStatic Ion Ring ExpEriment; DESIREE, at Stockholm University. We have produced beams of atomic carbon anions and small carbon anion molecules (C$_n^-$, $n=1,2,3,4$) in a sputter ion source. The ion beams were accelerated to 10 keV kinetic energy and stored in an electrostatic ion storage ring enclosed in a vacuum chamber at 13 K.
For 10 keV C$_2^-$ molecular anions we measure the residual-gas limited beam storage lifetime to be 448 s $\pm$ 18 s with two independent detector systems. Using the measured storage lifetimes we estimate that the residual gas pressure is in the 10$^{-14}$ mbar range.
When high current ion beams are injected, the number of stored particles does not follow a single exponential decay law as would be expected for stored particles lost solely due to electron detachment in collision with the residual-gas. Instead, we observe a faster initial decay rate, which we ascribe to the effect of the space charge of the ion beam on the storage capacity.
\end{abstract}


\maketitle
\section{Introduction}
\begin{figure}
	\centering
		\includegraphics[width=0.40\textwidth]{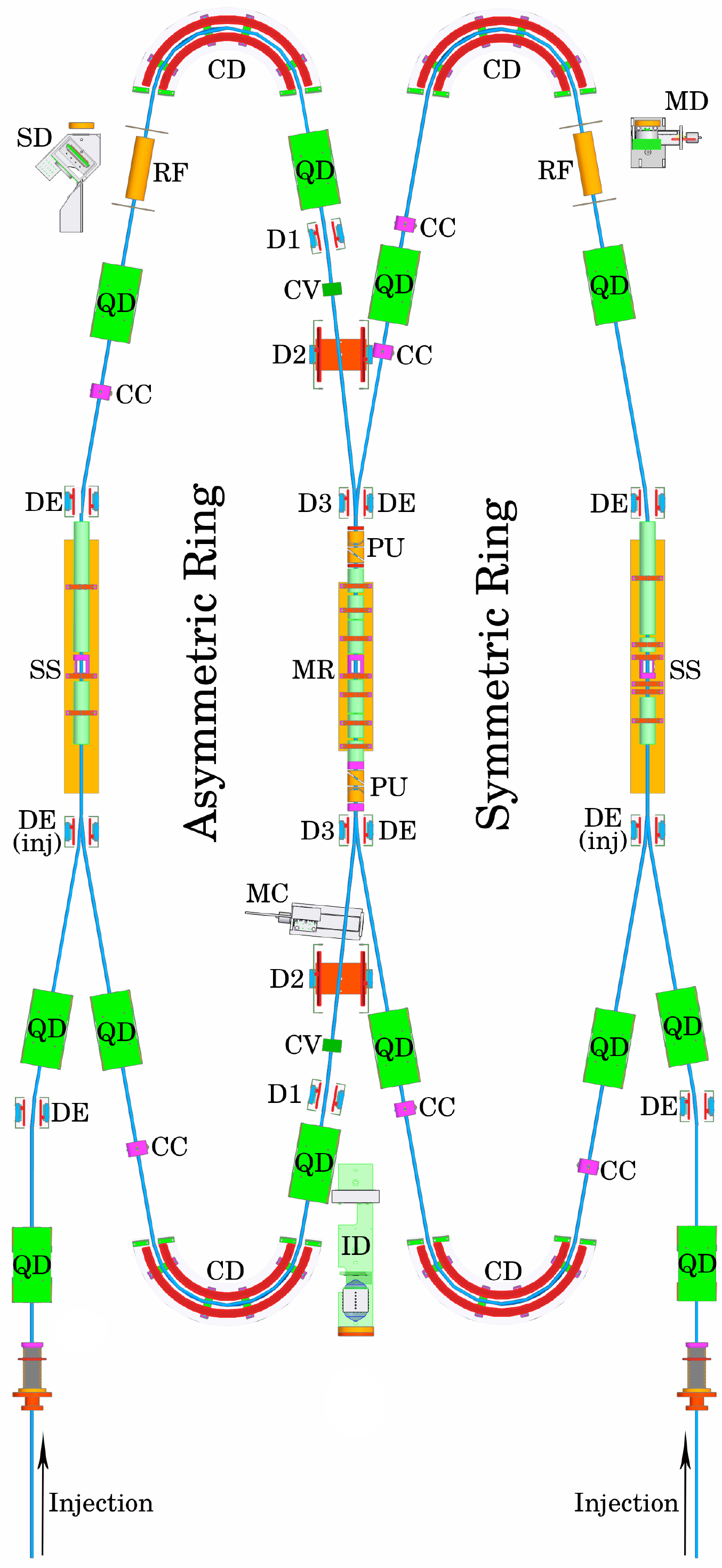}
	\caption{The electrostatic elements and detectors used for the two storage rings of DESIREE: CD- Cylindrical 160$^{\circ}$  Deflectors. DE- 10$^{\circ}$ Deflectors. QD- Quadrupole Doublets. MR- Merging Region enclosed in seven drift tube electrodes. SS- Straight Sections for collinear laser interaction. ID- Imaging Detector for neutral products from merging region. SD- Optically transparent secondary electron emission detector. MD-, MC- Movable position-sensitive detectors. D1-, D2-, D3- Beam-merging deflectors (note that the deflector plates denoted both D3 and DE bend the beam in the symmetric ring by 10$^{\circ}$ and the beam in the asymmetric ring by an angle determined by the two beam storage energies). CC-, CV- Correction elements. PU- Electrostatic pick-up electrodes. RF- Radiofrequency beam bunching electrodes.
	The parts of the injection lines included in this picture are situated inside the main DESIREE vacuum enclosure. Four position-sensitive detectors are mounted in DESIREE. Two of these (ID) and (MD) detect neutrals from the symmetric ring used in the present work. A more detailed technical description of the storage rings is given in Ref. \cite{Tho11}.}
	\label{fig:desireeoutline}
\end{figure}

The construction of the Double ElectroStatic Ion-Ring ExpEriment, DESIREE \cite{Tho11, Sch08Desiree} at Stockholm University is close to being fully completed. The primary objective of DESIREE is to study interactions between keV beams of positive and negative ions at very low and well controlled center-of-mass energies (down to a few meV) and at very low internal temperatures (down to a few K). To achieve this, two electrostatic ion-storage rings with circumferences $C$=8.6 m are constructed with a common merging section and mounted in the interior of a cryogenic aluminium vacuum chamber (see figure \ref{fig:desireeoutline} for the ion-optical layout). This vacuum chamber is situated inside an outer steel vacuum chamber and is cooled by the colder second stages of four two-stage cryogenerators. The first stages of the cryogenerators are thermally connected to a copper screen mounted inside the steel vacuum chamber. This screen encloses the cryogenic aluminium chamber completely with the exception of a small number of laser/viewing ports and the two ion-beam injection ports. For further insulation the copper screen is covered by 30 layers of superinsulation. With this system temperatures below 10 K may be reached. This low temperature also leads to very low outgassing rates from the vacuum chamber walls and therefore an excellent vacuum is achieved. Atomic-, molecular, and cluster-ion beams may then be stored for long times during which infrared active species will reach thermal equilibrium with the surroundings \cite{Tho11}. Here, we present results of the first cryogenic operation and ion storage experiments in DESIREE using beams of 10 keV atomic and small molecular carbon anions C$^-$, C$_2^-$, C$_3^-$, and C$_4^-$.

DESIREE will be unique in the sense that no other facility combines the merged-beams geometry with long-time storage in a cryogenic environment. However, there are a number of electrostatic storage devices that are in operation or under construction, which have some other features in common with DESIREE. The first electrostatic ion storage ring which was built is the ELISA ring (ELectrostatic Ion Storage ring in Aarhus) which has a circumference of 7 meters \cite{Mol97}. The ELISA project introduced the general layout of a closed orbit achieved with two cylindrical 160$^\circ$ bends and four 10$^\circ$ deflectors.
The same or very similar designs have been used for several of its followers.
One room temperature storage ring, similar to ELISA and equipped with a merged electron beam, was built in Tsukuba, Japan \cite{Tan03}. A liquid nitrogen cooled storage ring was constructed at the Tokyo Metropolitan University and has been in operation for a number of years \cite{Jin04}. A somewhat larger electrostatic storage ring with a very different ion-optical layout and which will be operated at room temperature \cite{Sti10} has been constructed at Frankfurt University. The Frankfurt ring is currently being commissioned as is a smaller 4 m circumference room temperature ring at Aarhus University \cite{lha13}. At the Max-Planck Institute for Nuclear Physics in Heidelberg, a very ambitious project is close to its realisation: A 35 m circumference single cryogenic electrostatic ion-storage ring (CSR) is under construction and will be equipped with an internal gas-jet target, an ion-neutral merging section, and an electron cooler/target device \cite{Hah11}. A single cryogenic ring with a circumference of about 3 meters is under construction at RIKEN in Japan \cite{Azu13}. Furthermore, a new electrostatic ring is being built at King Abdulaziz City for Science and Technology in Riyadh \cite{Gha13}.

In parallel with the development of the larger electrostatic storage rings there has been a development of much smaller and compact devices in the form of electrostatic ion-beam traps (EIBTs). In these traps, the ions are moving with keV kinetic energies between two focusing electrostatic mirrors. The first EIBTs were developed simultaneously and apparently with no mutual communication between the groups at the Weizmann Institute \cite{Zaj97} and at Berkeley \cite{Ben97}. The design of the Israeli trap is often referred to as the Zajfman trap and several laboratories have developed their own versions of this---generally room-temperature operated---device. One such trap has been built at Queen's University, Belfast \cite{Ale09}. As part of the CSR project in Heidelberg, a Zajfman trap has been installed in a cryogenic chamber and storage lifetimes of several minutes have been demonstrated \cite{Lan10}. Another simpler design of an EIBT is the so-called ConeTrap, which has been developed at Stockholm University \cite{Sch01}. This device has been installed in a cryogenic test chamber for the DESIREE project and a high-precision measurement of the lifetime of the metastable He$^-$ ion has been performed \cite{Rei09}. With this experiment, the large advantage with cryogenic ion storage was clearly demonstrated as the weakly bound (77 meV) He$^-$ ions were not destroyed by absorption of black-body radiation at an operating temperature of 10 K. 
New interesting developments are two room-temperature table-top electrostatic ion storage rings: The Miniring in Lyon \cite{Ber08}, which is in operation \cite{Mar13}, and the $\mu$-Ring at Tokyo Metropolitan University, which will be commissioned in 2013 \cite{Shi13}.

\section{Experiment}
Negative atomic and small molecular carbon ions are produced in a sputter ion source and accelerated to 10 keV. A $90^{\circ}$ double-focusing bending magnet is used to select a specific ion species, which is then transported by means of electrostatic steering and focusing elements to the DESIREE storage ring through several stages of differential pumping. A set of deflector plates in the beam line are connected to a high voltage switch to form short pulses of ions for injection in one of the DESIREE storage rings. In figure \ref{fig:desireeoutline}, the ion-optical elements of the two rings are shown schematically. The ring to the right has fourfold symmetry and is referred to as the symmetric ring.
This is where the lighter of two beams in a merged beams experiment, aiming at low ion-ion collision energies, will be stored. In order to steer and position the two ion beams for merging, additional horizontal steerers are needed in the other ring where the heavier particles are stored. Therefore the ring to the left in figure \ref{fig:desireeoutline} is referred to as the asymmetric ring. The test experiments described here are performed in the symmetric ring.

The ion bunches enter the DESIREE vacuum enclosure at the lower right corner of the picture in figure \ref{fig:desireeoutline}. At the time of the first passage through the $10^{\circ}$ injection plates the ion bunch is deflected to the right in order to enter the closed orbit of the storage ring. Before the bunch has made a full revolution, the voltages on these injection plates are switched so that the bunch is deflected to the left after a full turn and these deflectors are then part of the closed-orbit ion optics for storage. The ion bunch durations are set to values of 5-10 $\mu$s, which is shorter than the revolution times of the stored ions (21.6 $\mu$s for 10 keV C$^-$ ions). Electronic noise on the ion optical elements may in principle limit the ion storage time, especially at the very low residual gas pressures at which we are normally operating.
For this reason we use 5 Hz low-pass filters in the power supplies and 
in the HV amplifiers for all ion optical elements in the storage rings. The noise levels have been measured to be better than 1 ppm rms of the maximum voltages. For the injection plates, where fast rise and fall times (100 ns) are required, switched low pass-filters are coupled between the outputs of fast Behlke switches (type: HTS301-03-GSM) and the deflector electrodes to reduce the noise of the switches during ion storage. During injection, these switched low-pass filters are inactivated by HV relays for a few ms. The key here is that after a few ms of ion storage, the low-pass filters are activated and the noise is again damped by about 40 dB.
For detection of  stored ion-beam pulses, electrostatic pick-up electrodes and detectors for neutralized particles are available. Horizontal and vertical pick-up electrodes are placed at both ends of the merging section and are thus common to the two rings. Two detectors for neutrals are mounted at the symmetric ring. One is a triple-stack 75 mm diameter microchannel plate detector with a phosphor screen anode which is placed after the merging section (ID in figure \ref{fig:desireeoutline}). For the present test the phosphor screen is viewed from the outside by a photomultiplier tube, but a CCD camera is also available in order to obtain position information \cite{Ros07,Tho08}. The other detector, mounted after the straight section on the injection side of the ring, is a 40 mm triple-stack microchannel plate detector with a resistive anode (MD in figure \ref{fig:desireeoutline}). This detector can be moved by means of a cryo- and vacuum-compatible stepper motor to allow passage of collinear laser beams.

The actual storage tests are performed by injecting an ion bunch and then recording the count rate on the two neutral-particle detectors (ID and MD) as a function of time after the injection. A PC-based multi-channel scaler system is used to collect the data. In addition, the signals from electrostatic pick-up electrodes in the merging section are recorded. Beams of the atomic anion C$^-$ have been stored when the ring was operating at room temperature and at a pressure in the 10$^{-8}$ mbar range. Further, measurements have been performed at cryogenic conditions including storage of 10 keV C$^-$, C$_2^-$, C$_3^-$ and C$_4^-$ ion beams. For cryogenic operation the storage lifetimes much exceed the time scale for the de-bunching of the injected ion pulses and we could determine the relative velocity spread of an injected C$^-$ beam to be about 1$\times$10$^{-3}$ from the temporal development of the width of a short ion bunch after injection.

\section{Results and Discussion}
\begin{figure}
	\centering
		\includegraphics[width=0.50\textwidth]{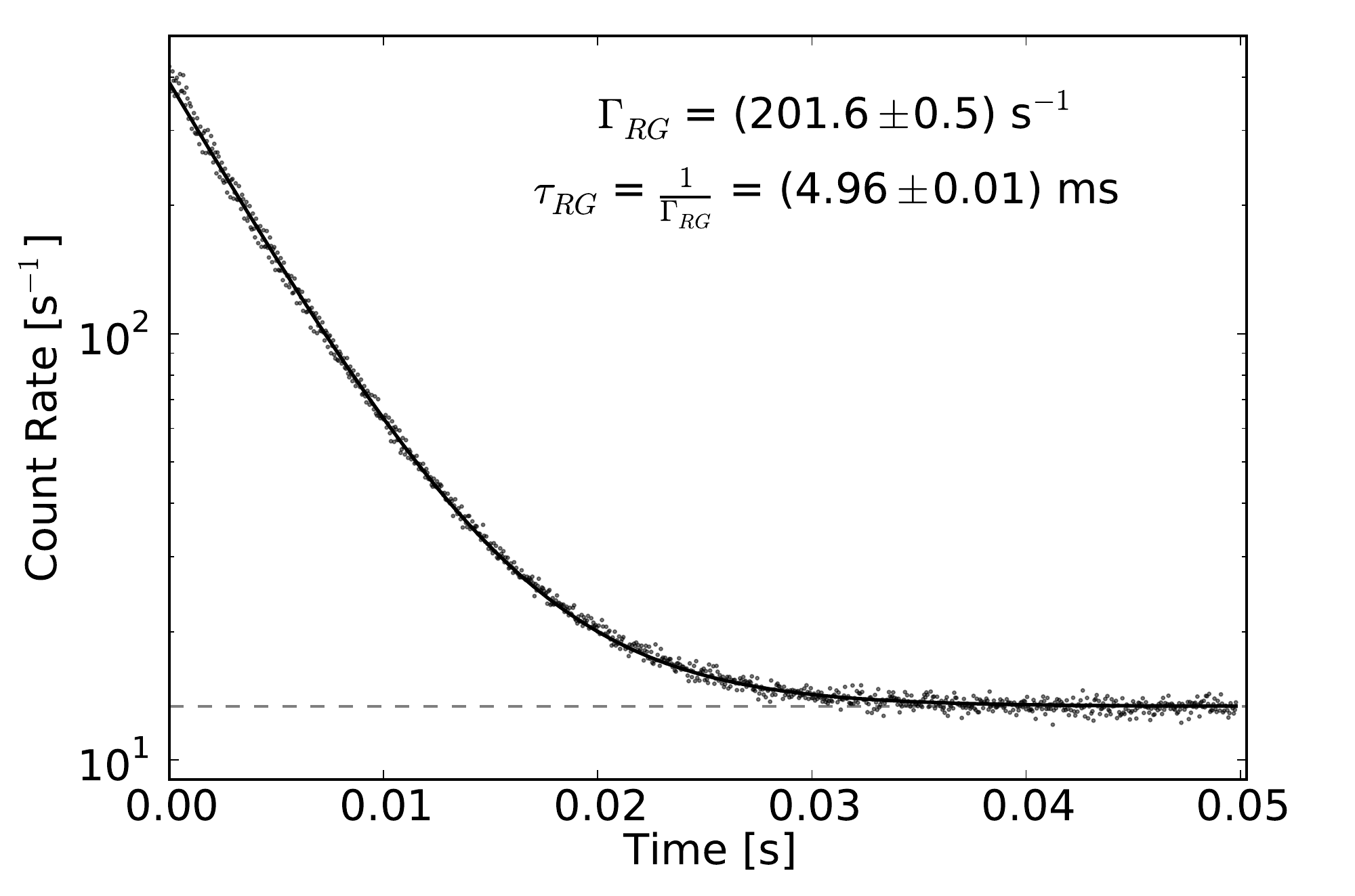}
	\caption{Count rate of neutral C atoms impinging on the imaging detector (ID in figure \ref{fig:desireeoutline}) as a function of time after injection of a 10 keV C$^-$ beam. The measurement was performed at room temperature and with a residual-gas pressure of $5.0 \times 10^{-8}$ mbar. The full curve is a fit to the data with a single exponential function plus a constant due to the detector background count rate without the ion beam.}
	\label{fig:decay_warm}
\end{figure}
In figure \ref{fig:decay_warm} the count rate of neutral C atoms detected by the imaging detector after the beams-merging section (ID, in figure \ref{fig:desireeoutline}) is shown as a function of the time after injection for 10 keV C$^-$ ions stored at 300 K. The data follows the functional form of a single exponential decay plus a low constant rate of detector dark counts. A numerical fit to a function of this form yields an average storage lifetime of $\tau$=5.0 ms with an insignificant statistical error (cf. figure \ref{fig:decay_warm}). During this test the vacuum gauge reading showed a residual-gas pressure of $p=5.0 \times 10^{-8}$ mbar.
We assume that electron detachment in residual-gas collisions is the dominating loss mechanism and deduce the average effective detachment cross section from the measured rate of destruction of the anions in residual-gas collisions $\Gamma_{RG}=1/\tau$:
\begin{equation}
\langle \sigma \rangle_{300K} = \frac{\Gamma_{RG}}{C \cdot f} \frac{k_B T}{p} = 4.2 \times 10^{-15} \text{ cm}^2,
\label{eqn:sigma}
\end{equation}
where $k_B$ is Boltzmann's constant, $f$=46.3 kHz is the revolution frequency of the stored ions, and $T$=300 K is the temperature of the vacuum chamber.
Luna {\it et al.} \cite{Lun01} found for C$^-$ at very similar velocity that the detachment cross sections were about $1 \times 10^{-15}$ cm$^2$ for noble gas targets. We ascribe our higher value to the larger physical dimensions of the molecules of the residual gas compared to the noble-gas atoms of ref. \cite{Lun01}.

\begin{figure}
	\centering
		\includegraphics[width=0.50\textwidth]{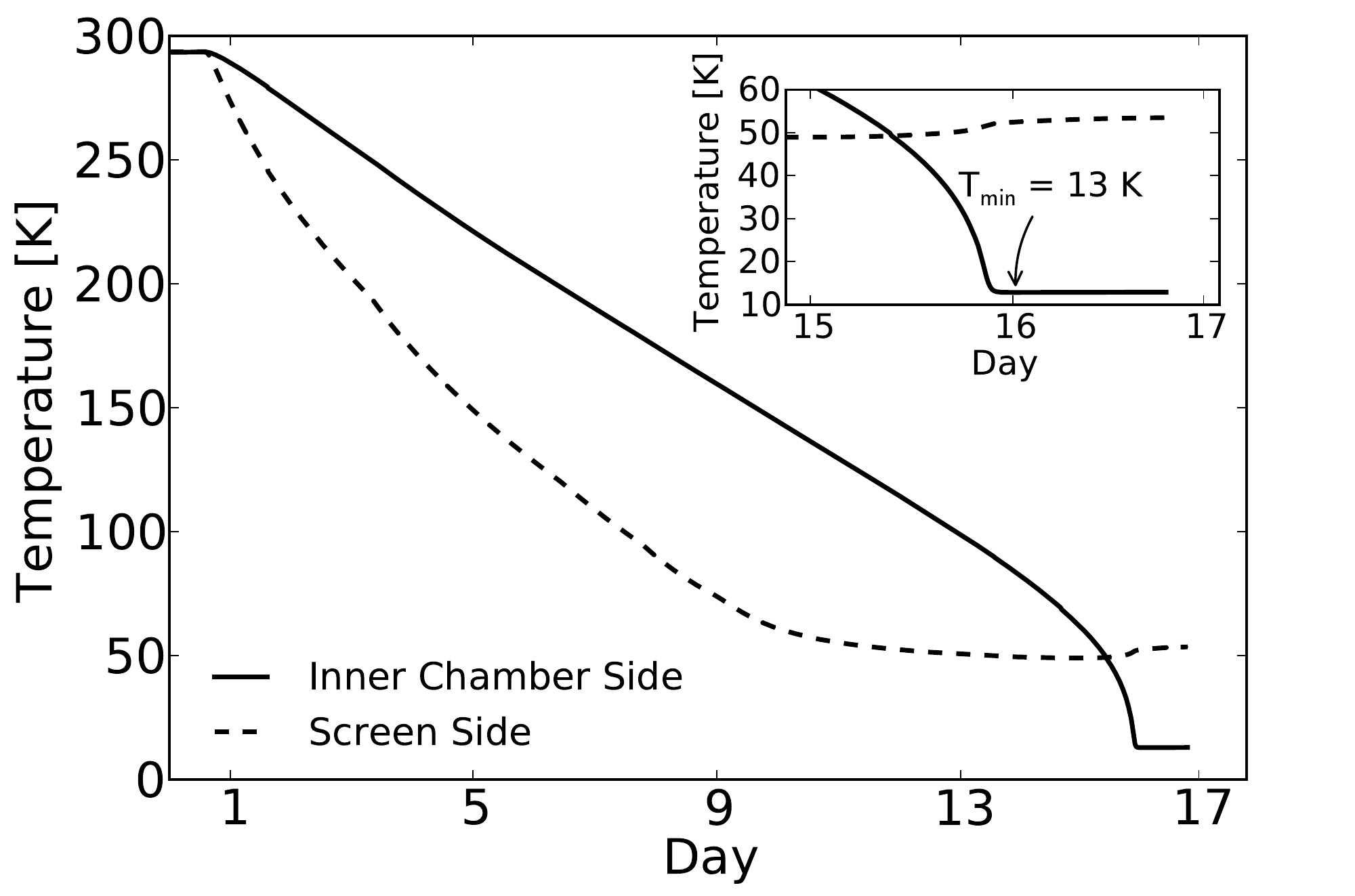}
	\caption{Measured temperatures of the inner vacuum chamber and of the thermal copper screen as functions of time after the cryocoolers were first switched on in September, 2012. The temperatures were monitored by temperature sensitive silicon diodes.}
	\label{fig:cooldown}
\end{figure}
For cryogenic operation the four closed-circuit He cryogenerators are used. In figure \ref{fig:cooldown} the temperatures of the inner vacuum chamber walls and the thermal screen are shown as a function of time after the cryogenerators were first switched on. All ion-optical elements are thermally connected to the inner chamber and will assume the same temperature. The final inner-chamber temperature of 13 K was reached after 16 days of continuous cooling. The rapid decrease in temperature on the last day of cooling down is due to the significant reduction of the heat capacity of the materials at temperatures below 100 K. In this initial cool-down experiment, the motion feedthrough connector to the Ti-sublimation pumps was unfortunately stuck which gave a substantial additional heat load to the system. This connector is designed to serve six Ti-sublimation pumps individually and normally it will be retracted such that there is no mechanical connection from the outside to the inner chamber. As this problem has now been solved we expect to reach still lower temperatures when the system is cooled next time. The results presented in the following discussion were, however, obtained at 13 K.

As all ion-optical elements are mounted on a common aluminum plate, it was expected that only small effects on the ion-storage conditions would result from changing the temperature. Indeed, it was found that the exact same ring settings yielded storage at both room and cryogenic temperatures. To reach optimum conditions, some fine tuning is, however, needed when changing between room and cryogenic temperature operation. Further it is necessary to modify the injection settings due to the relative displacement of the storage ring and the injection line caused by the cooling of the former.
\begin{figure}
	\centering
		\includegraphics[width=0.50\textwidth]{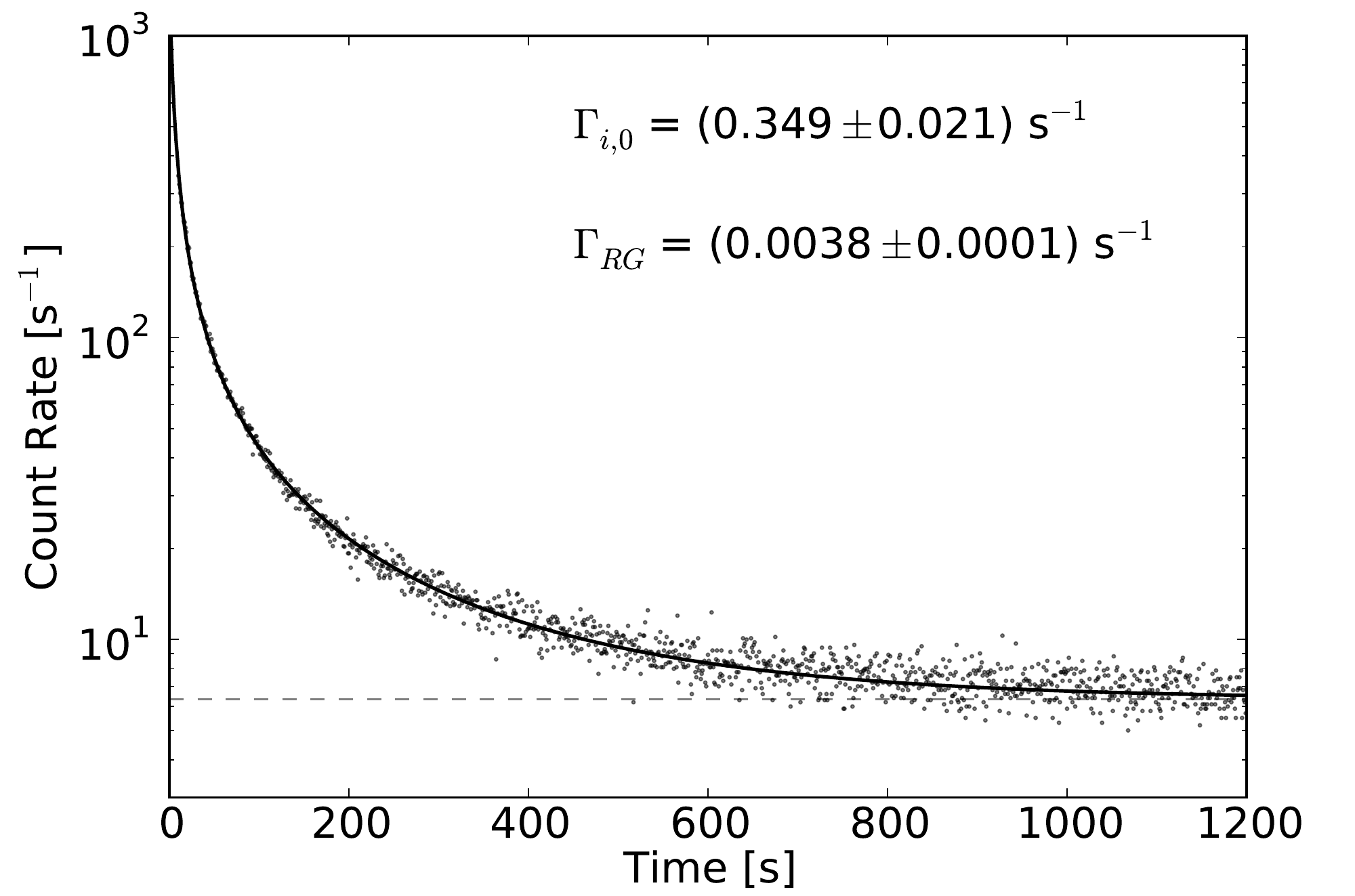}
	\caption{Count rate on the movable detector for neutral particles (MD in figure \ref{fig:desireeoutline}) as a function of time after the injection of a beam of 10 keV C$^-$ ions. The temperature of the inner vacuum chamber was $T$=13 K.}
	\label{fig:C-_cold01}
\end{figure}
In figure \ref{fig:C-_cold01} the rate of neutral C atoms is shown as a function of time after injection of a 10 keV C$^-$ beam at cryogenic operation. Some very dramatic differences compared to the corresponding room temperature results of figure \ref{fig:decay_warm} are evident. First of all the beam storage time is now much longer. At cryogenic temperatures the residual-gas density is much reduced and therefore the loss rate due to collisional electron detachment is similarly reduced. While there is no direct measurement of the pressure in the cryogenic environment, we will, in the following analysis, use the measured loss rate to infer the residual-gas density. A second important difference between cold and warm operation is that the count rate no longer follows the single-exponential behavior characteristic of a situation with a constant decay rate due to residual-gas collisions. Instead, the decay is faster initially when a large number of ions are circulating together. Thus, another loss mechanism is at play, which is caused by interactions between the ions themselves. This could be related to the space-charge of the ion beam counteracting the effects of the quadrupole doublets used to focus the beam and obtain stable closed orbits. 

To model the decay we follow the same approach as was succesfully applied to describe the population of ions in the small electrostatic ion-beam trap, ConeTrap \cite{Reinhed2010}.
We assume that the total loss rate, $\Gamma$, can be expressed as the sum of two contributions: $\Gamma_{RG}$, which is the constant rate of loss due to collisions against the residual gas ($RG$) and $\Gamma_{i}$, which is the rate of loss due to the space charge of the ion ($i$) beam itself. The latter is assumed to be proportional to the number of ions stored, $N(t)$, and the total loss rate is:
\begin{equation}
\Gamma(t) = \Gamma_{RG} + \Gamma_{i,0} {N(t) \over N_0}
\label{eqn:Gamma}
\end{equation}
where $N_0$ is the initial number of stored ions and $\Gamma_{i,0}=\Gamma_{i}(t=0)$ is the initial additional loss rate induced by the ion beam itself.
The differential equation for the number of stored ions as a function of time is then,
\begin{equation}\label{rate equation}
\frac{dN(t)}{dt}=-\Gamma_{RG} N(t)-\Gamma_{i,0}\frac{N(t)^2}{N_0}.
\end{equation}
The solution to (\ref{rate equation}) is:
\begin{equation}\label{solutionrateeq}
N(t)=\frac{N_0}{(1+\Gamma_{i,0}/\Gamma_{RG}) e^{t \Gamma_{RG}}-\Gamma_{i,0}/\Gamma_{RG}}.
\end{equation}

For the room temperature data shown in figure \ref{fig:decay_warm} we have a situation where $\Gamma_{RG} >> \Gamma_{i,0}$ and then equation (\ref{solutionrateeq}) reduces to a single exponential decay as measured. Note also that the exponential growth in the denominator implies that after some time the expression (\ref{solutionrateeq}) reduces to a single exponential irrespective of the value of the ratio, $\Gamma_{i,0}/\Gamma_{RG}$. This is to be expected since the importance of the space charge must reduce as the beam intensity decreases.

The full curve in figure \ref{fig:C-_cold01} is the result of a fit to the data with a function of the form (\ref{solutionrateeq}) plus a constant detector background, which is measured separately without ions in the storage ring.
The three fit parameters are the initial count rate at $t=0$, $R(t=0)$, $\Gamma_{RG}$, and $\Gamma_{i,0}$, which is proportional to $N_0$.
A very good fit to the data is found with a reduced chi-square of $\chi^2=1.04$. The results are $\Gamma_{i,0}=0.349(21)$ s$^{-1}$ for the initial ion-induced loss rate and $\Gamma_{RG}=0.0038(1)$ s$^{-1}$ for the constant loss rate ascribed to residual-gas collisions. The latter corresponds to a 1/e lifetime of $\tau=1/\Gamma_{RG}=265(5)$ s. No instrument capable of measuring the pressure in the cold inner chamber directly is available today.
With the determination of $\tau$ as outlined here, we can, however, determine an approximate value of the pressure under the somewhat bold assumption that the effective cross section for destruction is the same for the residual-gas compositions at 300 K and 13 K.
Assuming the same average destruction cross section, we
find that at the time of acquisition of the data set shown in figure \ref{fig:C-_cold01}, the residual-gas density was
$n=2 \times 10^4$ cm$^{-3}$
corresponding to a pressure of
$p=4 \times 10^{-14}$
mbar at 13 K.
The uncertainties in the values for the density and pressure are estimated to be about a factor of two, which includes the uncertainty of the room temperature pressure measurement.

\begin{figure}
	\centering
		\includegraphics[width=0.50\textwidth]{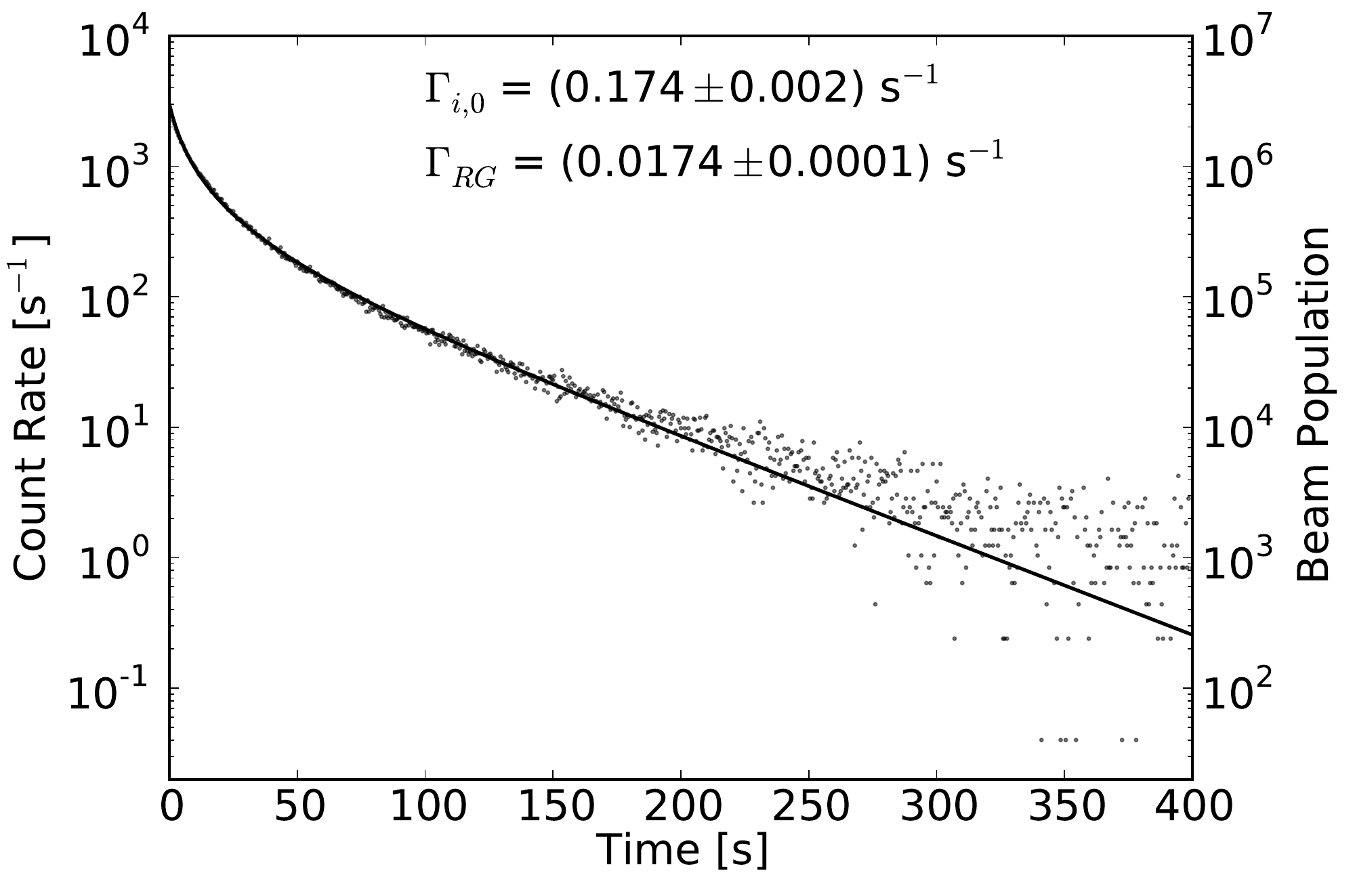}
	\caption{Count rate of neutrals as a function of time after the injection of an intense beam of 10 keV C$^-$ ions with the inner vacuum chamber at 13 K. The constant rate of detector dark counts has been subtracted and the vertical axis on the right gives the number of stored ions (cf. text).}
	\label{fig:C-_cold02}
\end{figure}


We may deduce the number of stored ions at time $t$, $N(t)$, from the count rate, $R(t)$ {\it without making any assumptions} regarding the residual-gas density and composition and the related electron detachment cross sections. Given that the process producing neutral particles is collisional detachment in the residual gas, the count rate, $R(t)$, is given by:
\begin{equation}\label{count rate}
R(t)=L \langle n\sigma \rangle f\epsilon N(t),
\end{equation}
where $L=$95 cm, is the length of the straight section from which neutralised projectiles will strike the detector and $\epsilon$ is the detection efficiency, which for the resistive anode encoder detector has a maximum value of $\epsilon$=0.50. The quantity $\langle n\sigma \rangle$ is the sum of the products of densities and electron detachment cross sections for the residual-gas components. Note that this relation holds even when the collisional detachment is not the main loss mechanism of particles from the beam as long as these other losses do not lead to the detection of neutrals. 

In fact, $\langle n\sigma \rangle$ is the reciprocal of the average distance travelled by an ion before electron detachment. It is thus related to the measured decay rate due to residual-gas collisions, $\Gamma_{RG}$, through:
\begin{equation}\label{nsigma}
\langle n\sigma \rangle=\frac{\Gamma_{RG}}{Cf}.
\end{equation}
Using this in equation (\ref{count rate}) we find:
\begin{equation}\label{number of ions}
N(t)=\frac{C}{L} \frac{1}{\epsilon} \frac{1}{\Gamma_{RG}} R(t).
\end{equation}

In figure \ref{fig:C-_cold02} we show a data set for 10 keV C$^-$ ions stored under somewhat different conditions than the set plotted in figure \ref{fig:C-_cold01}. For the figure \ref{fig:C-_cold02} test run the vacuum was not quite as good as for the data in figure \ref{fig:C-_cold01}, but the intensity of the ion beam was higher. We have applied the relation in equation (\ref{number of ions}) and subtracted the constant detector background to deduce the number of stored ions, $N(t)$, as a function of time. Initially the number of stored anions was around $N_0 = 3.0 \times 10^6$ corresponding to an ion current of 22 nA  under the assumption that the detection efficiency had its maximum value of $\epsilon$=0.5.
Thus, 22 nA is a lower limit for the initially stored ion current yielding the data in figure \ref{fig:C-_cold02}.

Electrostatic storage of 20 keV C$^-$ ions at temperatures between 77 K and room temperature has been studied earlier by Takao {\it et al.} \cite{Tak07}. Their room temperature storage lifetime result is consistent with ours when the different vacua and beam velocities are considered. Further, they measured the lifetime of a metastable state as function of temperature. This state lies close to the continuum and is therefore sensitive to photodetachment by thermal radiation. In the present study, we do not observe any clear evidence of this metastable state, neither at room temperature (due to the modest vacuum conditions at which the present room temperature data set was recorded) nor at 13 K (due to the absence of photodetaching thermal photons).

\begin{figure}
	\centering
		\includegraphics[width=0.50\textwidth]{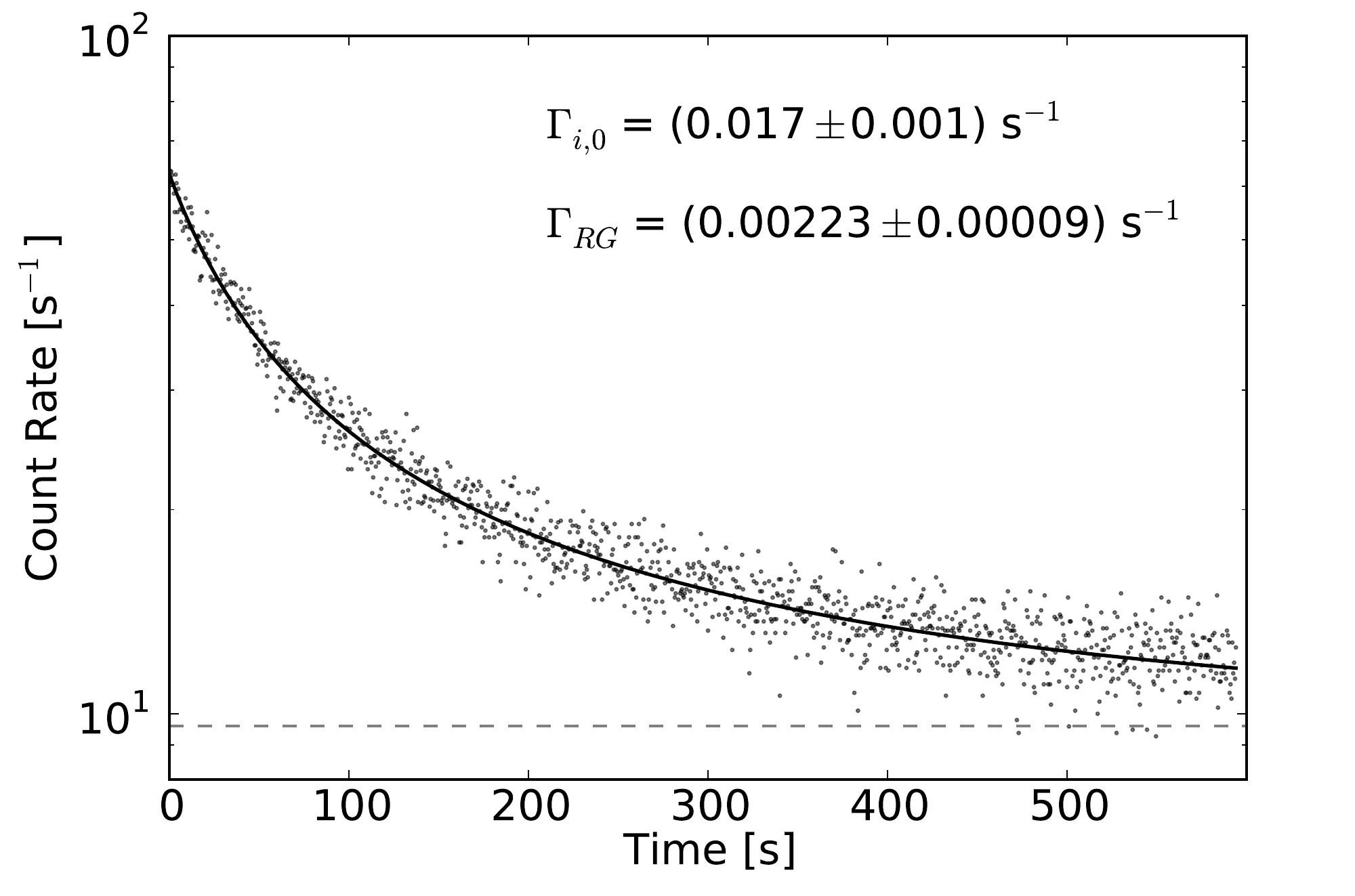}
	\caption{Count rate of neutrals on the imaging detector (ID in figure \ref{fig:desireeoutline}) as a function of time after the injection of a beam of 10 keV C$_2^-$ ions with the inner vacuum chamber at 13 K.}
	\label{fig:C2-_cold}
\end{figure}

Figure \ref{fig:C2-_cold} shows the result of a storage measurement for a 10 keV C$_2^-$ molecular ion beam and a fit to equation (\ref{solutionrateeq}). We deduce a residual-gas limited storage lifetime of $\tau=1/\Gamma_{RG}=(448 \pm 18)$ s, which is the longest DESIREE storage lifetime measured so far.

\section{Conclusions and outlook}
One of the two storage rings of the Double ElectroStatic Ion-Ring ExpEriment, DESIREE, has been commissioned. Beams of C$^-$, C$_2^-$, C$_3^-$ and C$_4^-$ ions have been accelerated to 10 keV and stored in the electrostatic storage ring with residual-gas collision limited lifetimes of up to 7.5 minutes at T=13 K.
Residual-gas pressures in the 10$^{-14}$ mbar range were deduced from the observed storage lifetimes. Initial non-exponential losses are ascribed to ion-beam space-charge effects. The number of stored ions follow a functional form corresponding to a decay rate with two terms: One constant term describing losses through collisional electron detachment with the residual-gas and one term, which is linear in the number of stored particles. Using the constant term, the proportionality factor between the detector count rate and the number of stored particles is deduced. In this way a {\it lower} limit of the initially stored ion current of 22 nA was found for a storage test with 10 keV C$^-$ ions.

Two minor technical issues were not solved at the time of the collection of the data presented here. As mentioned earlier a significant additional thermal load was present due to a stuck linear motion feedthrough. We therefore expect that a lower temperature will be reached at the next cool-down, which may further lead to a lower pressure due to a significantly improved pumping of H$_2$, which is the dominating component of the residual gas. Additionally, a small leak directly from atmosphere into the inner vacuum chamber has been located and repaired. This should definitely also influence the vacuum on the level of 10$^{-14}$ mbar.

With the results presented here, we are now ready to perform single-ring experiments at cryogenic temperatures. This means that we can make improved experiments with for examples loosely bound systems such as C$_{60}^{2-}$ \cite{Tom06} 
and study inherent stabilities of C$_n^{2-}$ dianions (Cs charge-exchange cells are available on the DESIREE injection beam lines).
Furthermore, cooling of stored molecules \cite{Mar13,Got13} can be followed to very low temperatures. In the near future the asymmetric ring will be
commisioned. This will make the facility fully operational 
enabling event-by-event studies of interactions between internally cold positive and negative ions at meV center-of-mass collision energies.
\begin{acknowledgments}
We are thankful for the investment grants from the Knut and Alice Wallenberg Foundation and the Swedish Research Council and for the operation grant from the Swedish Research Council (contract 2011-6314). HTS is grateful to the Swedish Research Council for financial support under contract 2011-4047. RDT is grateful for the support of the European Office of Aerospace Research and Development for this work under Project FA8655-11-1-3045. HC gratefully acknowledges the support from the Swedish Research Council under contract 2012-3662.
\end{acknowledgments}


\end{document}